\def\breakon{\end{multicols}\widetext\vspace{-.2cm}
\noindent\rule{.48\linewidth}{.3mm}\rule{.3mm}{.3cm}\vspace{.0cm}}
\def\breakoff{\vspace{-.2cm}
\noindent
\rule{.52\linewidth}{.0mm}\rule[-.27cm]{.3mm}{.3cm}\rule{.48\linewidth}{.3mm}
\vspace{-.3cm}
\begin{multicols}{2}
\narrowtext}
\documentstyle[aps,prl,multicol,epsfig]{revtex}

\def\beq{\begin{equation}}
\def\enq{\end{equation}}
\def\bqa{\begin{eqnarray}}
\def\eqa{\end{eqnarray}}

\begin{document}

\draft

\widetext

\title{An Adiabatic Quantum Pump of Spin Polarized Current}

\author{Eduardo R. Mucciolo$^1$, Claudio Chamon$^2$, and Charles
M. Marcus$^3$}

\address{
$^1$Departamento de F\a'{\i}sica, Pontif\a'{\i}cia Universidade
Cat\'olica do Rio de Janeiro, \\
C.~P. 38071, 22452-970 Rio de Janeiro, Brazil \\
$^2$Department of Physics, Boston University, Boston, MA 02215 \\
$^3$Lyman Laboratory of Physics, Harvard University, Cambridge, MA
02138}

\date{December 7, 2001}

\maketitle

 
\begin{abstract} 
We propose a mechanism by which an open quantum dot driven by two ac
(radio frequency) gate voltages in the presence of a moderate in-plane
magnetic field generates a spin polarized, phase coherent dc
current. The idea combines adiabatic, non-quantized (but coherent)
pumping through periodically modulated external parameters and the
strong fluctuations of the electron wave function existent in chaotic
cavities. We estimate that the spin polarization of the current can be
observed for temperatures and Zeeman splitting energies of the order
of the single-particle mean level spacing.
\end{abstract}


%


\begin{multicols}{2}

\narrowtext

The advent of shape-modulated quantum dots has renewed interest in the
problem of phase-coherent pumping of electrical charge by periodic
modulation of external parameters\cite{altshuler99}. The original idea
of coherent charge pumping, devised for gaped, isolated systems
\cite{thouless83}, has been extended to open systems
\cite{spivak95,brouwer98,zhou99}, and recently realized experimentally
by Switkes {\it et al.} \cite{switkes99}. Subsequent theoretical work
has focused on issues of symmetry, statistics, and phase coherence
\cite{shutenko00,aleiner00,vavilov00}, including a reinterpretation of
the experiment \cite{switkes99} as a rectification effect
\cite{brouwer00}. To date, there has been little discussion of
electron spin in quantum pumps.

There is a growing interest in the mesoscopic physics of spin
transport in microelectronic circuits \cite{prinz98}. Most coherent
spin transport devices proposed or realized experimentally so far are
based on the injection of polarized electrons from metallic or
semiconductor ferromagnetic contacts (for a recent review, see
Ref. \onlinecite{dassarma01}). An alternative approach based on
pumping of spin in purely one dimensional systems using fluctuating
gate voltages and magnetic fields has been recently proposed
\onlinecite{sharma01}.

In this Letter, we propose and analyze a new method of generating spin
polarized dc currents in semiconductors based on the parametric
pumping of spin without relying on spin injection. The basic idea is
to apply two cyclically oscillating gate voltages to a quantum dot
formed from a two-dimensional electron gas (2DEG) (similar to
adiabatic charge pumping) in the presence of a uniform magnetic field
applied in the plane of the 2DEG. The lifting of spin degeneracy by
the magnetic field allows an arbitrary ratio of spin and charge to be
pumped, including the situation in which a spin current of order
$\hbar$ per pumping cycle is produced with zero charge pumping. We
analyze the average amplitude of the spin polarized current for
experimentally realizable situations using perturbation theory. We
also briefly discuss important issues of dephasing, spin-orbit
coupling, and rectification effects.


The device we have in mind is an open quantum dot made from a confined
2DEG, with two point-contact leads connecting the dot to electron
reservoirs. The confining potential of the dot undergoes a periodic
shape deformation controlled by two ac gate voltages, $V_1(t) = A_1
\cos(\omega t + \phi_1)$ and $V_2(t) = A_2 \cos(\omega t + \phi_2)$
\cite{switkes99}, as shown schematically in Fig.  1(a). We assume that
the shape deformation is adiabatic, by which we mean $\omega \ll
\gamma_{\rm esc}$, where $\gamma_{\rm esc} = N \Delta/2\pi\hbar$ is
the escape rate from the dot, $\Delta = 2 \pi \hbar^2/m^\ast A$ is the
quantum level spacing of the closed dot with area $A$, and $N = N_r +
N_l$ is the total number of channels connecting the dot to the left
and right reservoirs.  We further assume for the sake of simplicity
that spin scattering, spin-orbit effects, and decoherence are
negligible, though in practice the latter two effects may have
significant consequences.


\begin{figure} 
\hspace{0.5cm}
\epsfxsize=7cm
\vspace{0.5cm}
\epsfbox{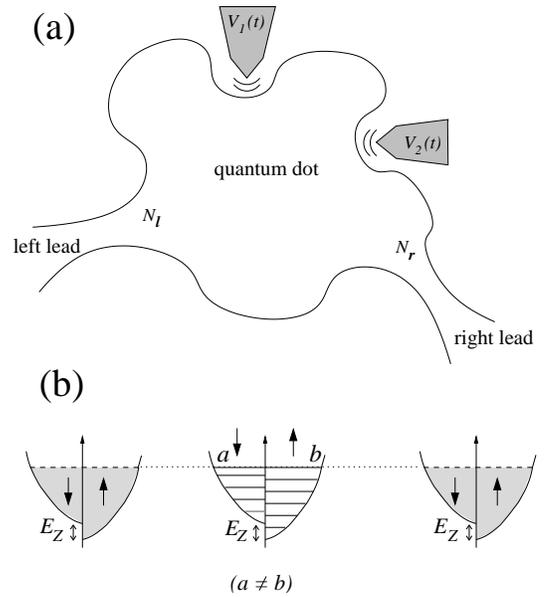}
\caption{(a). Schematic view of the device. (b). The energy level
diagram for the device in the presence of a parallel magnetic field.}
\label{fig:f1}
\end{figure}


\breakon

\begin{figure} 
\hspace{1cm}
\epsfxsize=14cm
\vspace{0.5cm}
\epsfbox{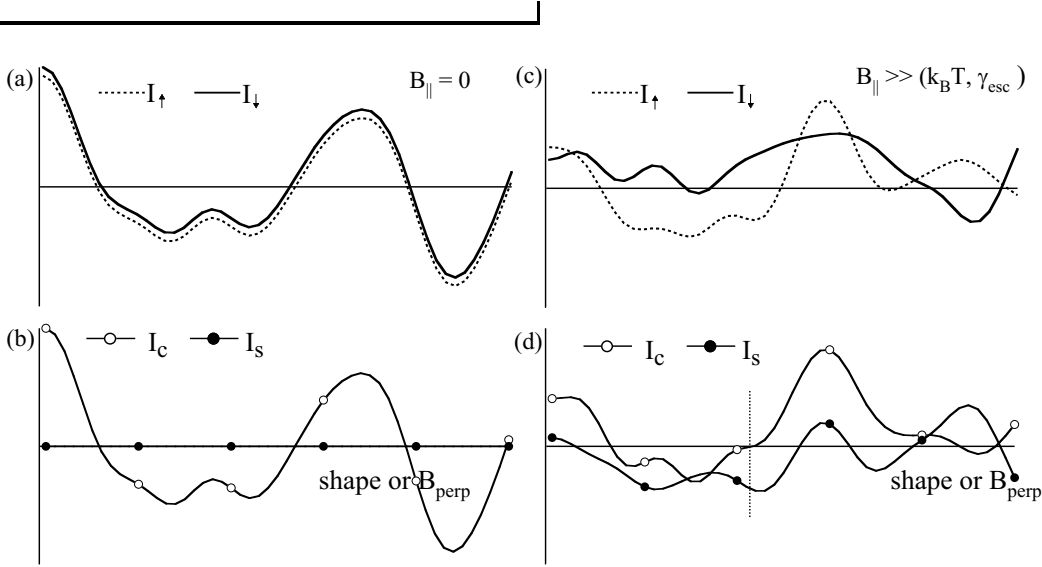}
 \caption{Pumped current dependence with an external parameter, such
 as dot deformation or perpendicular magnetic field. (a,b) No in-plane
 magnetic field $B_\parallel$ is applied. The spin up $I_\uparrow$ and
 spin down $I_\downarrow$ components of the charge current are equal
 and no spin current $I_s$ is present. (c,d) When a sufficiently
 strong in-plane field is applied, the spin component of the charge
 current respond in distinct ways to an external perturbation. As a
 result, total charge and total spin currents in the dot become
 uncorrelated. The dashed line in (d) indicates a point where only
 spin is transfered across the dot with no net charge transport.}
\label{fig:f2}
\end{figure}

\breakoff

In the absence of applied magnetic fields, the pumped current produced
by cyclic shape deformation of the dot carries no net spin, i.e., the
up and down spin components of the pumped current, $I_\uparrow$ and
$I_\downarrow$, are identical. In this case, $I_\uparrow$ and
$I_\downarrow$ fluctuate {\em together}, with zero average, as a
function of external parameters such as static dot shape and
perpendicular magnetic field. Spin degeneracy can be lifted by
applying a magnetic field in the plane of the 2DEG \cite{folk01}. For
moderate parallel fields, $E_Z = g^\ast\mu_B B_\parallel > (\hbar
\gamma_{\rm esc}, k_BT)$ (typically a few tesla for a micron scale GaAs
quantum dot at temperatures below 0.5K) the pumped currents associated
with the two spin directions $I_\uparrow$ and $I_\downarrow$ will
become uncorrelated, and will fluctuate independently as device
parameters are changed.

Let us denote the charge and spin currents passing through the dot as
$I_c$ and $I_s$, respectively: $I_{c,s} = I_\uparrow \pm I_\downarrow$
(we define spin current to have the same units of charge current,
understanding that $e \leftarrow \!\! \rightarrow \hbar/2$). Upon
averaging over different realizations of the dot shape or chemical
potential, $\overline{I_\uparrow} = \overline{I_\downarrow} = 0$. The
strength of the pumping current is characterized by its variance,
\beq
\overline{I_{c,s}^2} = \overline{I_\uparrow^2} +
\overline{I_\downarrow^2} \pm 2 \overline{I_\uparrow\, I_\downarrow} = 2
\left( \overline{I_\uparrow^2} \pm \overline{I_\uparrow\, I_\downarrow}
\right), 
\enq
where we assumed $\overline{I_\uparrow^2} =
\overline{I_\downarrow^2}$. In the absence of an in-plane field,
$\overline{I_\uparrow\, I_\downarrow} = \overline{I_\uparrow^2}$,
whereas for a strong enough applied field, we expect that incoming
spin up and down electrons will occupy uncorrelated sets of states in
the dot, leading to $\overline{I_\uparrow\, I_\downarrow} = 0$.  As a
result, $\overline{I_c^2}$ decreases by a factor of two in the large
field limit\cite{obs0}, while simultaneously, $\overline{I_s^2}$ goes
from zero to its maximum value. The most striking situation, however,
occurs when parameters are set such that $I_\uparrow = -
I_\downarrow$. In this case, a finite spin current exists through the
quantum dot without any net charge transport.  We note that it is
straightforward experimentally to tune parameters to achieve the
condition $I_\uparrow = - I_\downarrow$. Because $I_c$ fluctuates
randomly about zero as a function of external parameters, one can
simply tune dot shape or perpendicular field until the condition $I_c
= 0$ is found. This will be the state where $I_\uparrow = -
I_\downarrow$. This is illustrated in Fig. 2.


Let us call $Q_{\uparrow,\downarrow}$ the spin up/down charge
transfered after the completion of one cycle,
\beq
\label{eq:pumpedcharge}
 Q_{\uparrow,\downarrow} = \int_0^{2\pi/\omega} dt\,
 I_{\uparrow,\downarrow}(t).
\enq
For a chaotic or disordered quantum dot connected to leads with many
propagating channels ($N \gg 1$), the variance of pumped charge over
an ensemble of equivalent dots (e.g., differing in shape or disorder
configuration) has been calculated by Vavilov, Ambegaokar, and Aleiner
\cite{vavilov00}. We can generalize these calculations to include a
Zeeman field \cite{obs1}. For our purposes, it will be sufficient to
consider the theory in the limit of high temperature, when $\hbar
\omega \ll E_Z, k_B T, \hbar \gamma_{\rm esc}$ \cite{obs2}. The
resulting analytical expression for $Q_{\uparrow,\downarrow}$ is
further simplified if we restrict our analysis to the case of small
external oscillating voltages. This allows us to use an expansion in
powers of $A_1$ and $A_2$ and retain only the leading bilinear
term. Following Ref. \onlinecite{vavilov00}, we obtain
\bqa 
\label{eq:initial}
 \overline{Q_\uparrow Q_\downarrow} & = & \frac{16 \pi\ e^2 g\, C_1\,
 C_2\, \sin^2 \phi}{N\Delta} \int_0^\infty d\tau\, e^{-N
 \tau\Delta/\pi} \nonumber \\ & & \times (1 + N\tau\Delta/\pi)\,
 [F(\tau)]^2 \cos(2E_Z\tau),
\eqa
where $g = N_r N_l/N$, $\phi = \phi_1-\phi_2$, $F(\tau)=T\tau
/\sinh(2\pi T\tau)$, (we take $\hbar = k_B = 1$ hereafter). The
factors $C_{1,2}$ are related to the quantum dot response to shape
deformations and can be determined through their relation to the
quantum dot energy level susceptibility \cite{shutenko00,vavilov00}.

When the Zeeman energy is set equal to zero, Eq.  (\ref{eq:initial})
coincides with a similar expression in Ref.\ \onlinecite{vavilov00}
for $\overline{Q^2}$ and spinless electrons. Since $N \gg 1$, the
exponential factor dominates the integrand decay in
Eq. (\ref{eq:initial}) at low temperatures. In that case, the variance
of total spin transfered per cycle, $Q_s = Q_\uparrow - Q_\downarrow$,
will depend strongly on $N$.

The integral over $\tau$ can be evaluated numerically, yielding
results such as those shown in Fig. \ref{fig:f3}, where we have
plotted the quantity $r_{\rm pol} = \overline{Q_s^2}/
\overline{Q_c^2}$ versus $E_z$ for several values of $T$ and $N$, with
$Q_c = Q_\uparrow + Q_\downarrow$ and $\phi \neq 0,\pi$. Notice that,
at $E_Z = 0$, $\overline{ Q_\uparrow Q_\downarrow} =
\overline{Q_\uparrow Q_\uparrow}$, thus $r_{\rm pol} = 0$. As $E_Z$
grows, the amounts of up and down spin transfered per cycle become
uncorrelated. The typical amplitude of spin transfer depends strongly
on temperature. The dependence on $N$, which is pronounced at low
temperatures, decreases substantially when $T$ is of order $\Delta$
[see Fig. \ref{fig:f3} (b)].

From Eq. (\ref{eq:initial}) we can estimate the typical Zeeman energy
$E_Z^\ast$ necessary to achieve $r_{\rm pol} = 1/2$, {\it i.e.}, that
spin polarize $\sqrt{1/2}\approx 70\%$ of the pumped current.  When $T
\ll 2\pi \gamma_{\rm esc}$, we obtain $E_Z^\ast \approx 1.17\,
\gamma_{\rm esc}$, while in the opposite limit, $E_Z^\ast \approx
1.49\, T$. For a GaAs quantum dot with $1\, \mu$m$^2$ in area at 100
mK and 1 tesla, we find that the pumped current is typically 60\% spin
polarized ($r_{\rm pol} = 0.36$) when the total number of propagating
channels in the leads is 4.

Spin-flip scattering limits the efficiency of the spin current
pump. While several mechanisms could cause spin flipping, perhaps the
most relevant one to semiconductor materials is spin-orbit coupling
caused by asymetries in the confining potential and lattice
structure. In a small quantum dot at $B_\parallel = 0$, there is a
substantial reduction of the spin-orbit scattering rate as compared to
the bulk two-dimensional electron gas in a GaAs heterostructure
\cite{nazarov00,halperin01}. However, it is also know that the
presence of an in-plane magnetic field (such as the one needed for the
operation of the spin pump) alters significantly weak localization
corrections of the conductance in laterally confined quantum
dots\cite{folk01,halperin01,falko01,brouwer01,aleiner01}, suggesting
an enhancement of spin-orbit effects at $B_\parallel > 0$. This
enhancement does depends strongly on the size of the quantum dot, as
observed experimentally by Folk {\it et al.} \cite{folk01} and
theoretically examined by Halperin {\it et al.} \cite{halperin01}. For
example, for the dots in Ref. \cite{folk01}, there is a crossover to
strong spin-orbit coupling for large dots (8$\mu$m$^2$ in area), while
no substantial spin-orbit effects are detected for smaller dots
(1$\mu$m$^2$ in area). These results suggest that for small quantum
dots, in the regime of temperatures and Zeeman energies that we
discussed above in our estimation for 1$\mu$m$^2$ dots, spin-orbit
scattering should not be sufficient to destroy the spin pumping
mechanism we propose.


\begin{figure} 
\hspace{0cm}
\epsfxsize=7cm
\vspace{-.7cm}
\epsfbox{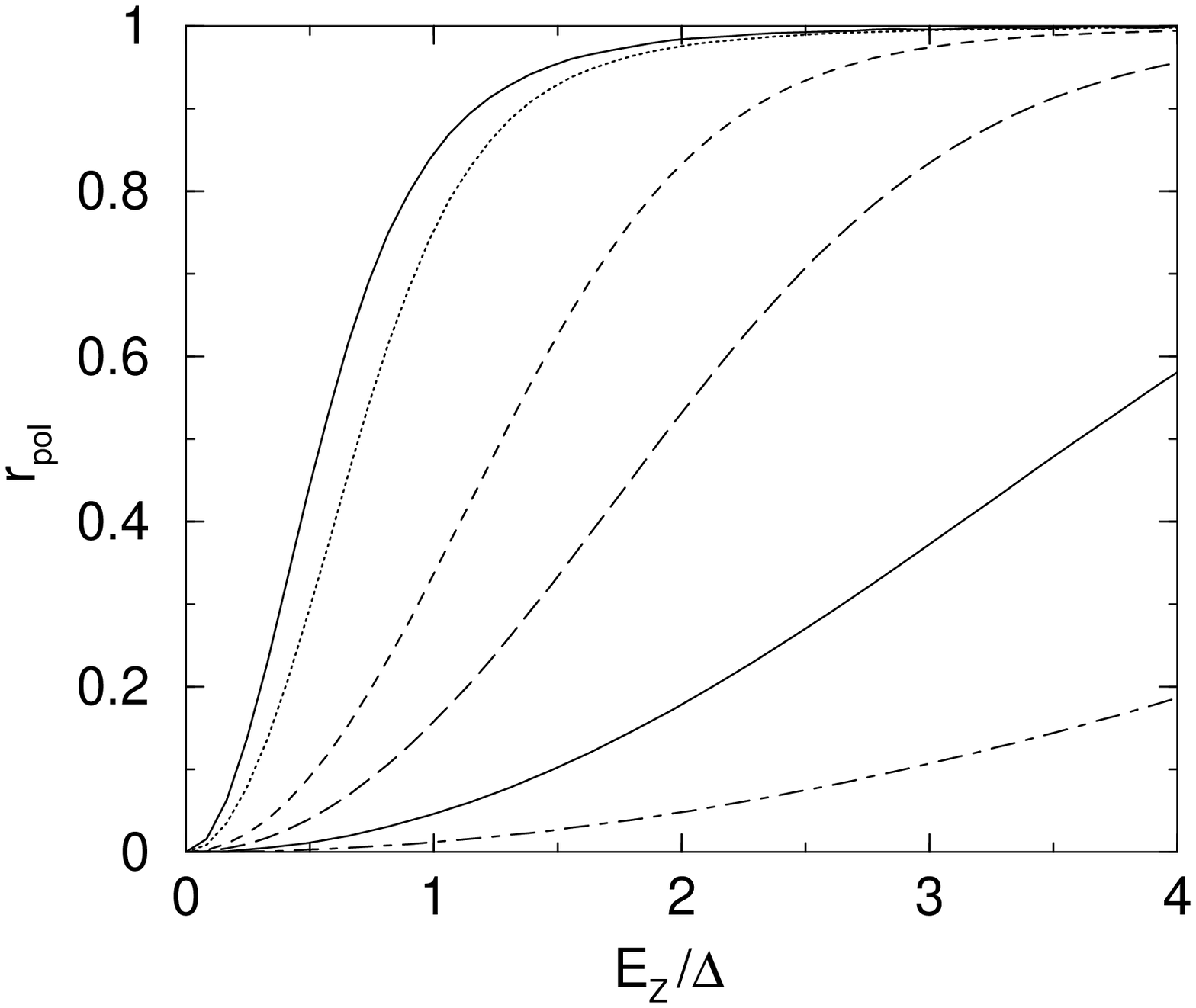}
\end{figure}
\begin{figure} 
\hspace{0cm}
\epsfxsize=7cm
\vspace{0cm}
\epsfbox{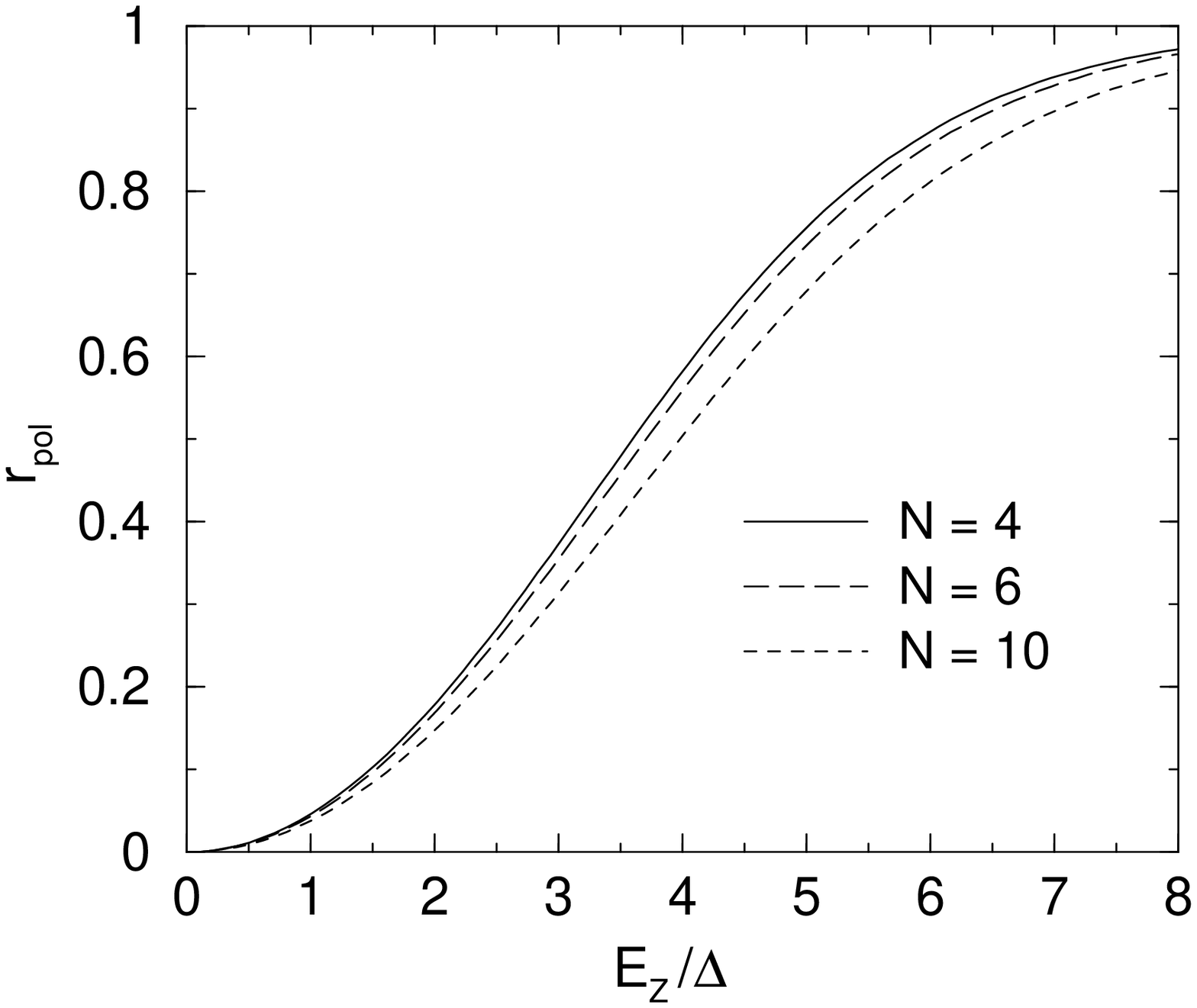}
 \caption{Relative spin polarization of the pumped current as a
function of Zeeman energy for: (a) $N=4$ and different temperatures
($T/\Delta = 0$, $0.1$, $0.3$, $0.5$, $1.0$, and $2.0$); (b)
$T=\Delta$ and different numbers of channel.}
\label{fig:f3}
\end{figure}


Another relevant question to be considered is whether the dc current
spin polarization effect caused by pumping in the presence of an
in-plane Zeeman field could be also generated by a rectification
mechanism \cite{brouwer00}. The answer is positive, since spin
polarization also appears when there is a difference between the
quantum dot charge conductance for up and down spin channels. That is,
provided $G_\uparrow(t)$ and $G_\downarrow(t)$ oscillate with distinct
amplitudes, for a voltage drop $V(t)$ we would have
$\overline{I_\uparrow(t)} \neq \overline{I_\uparrow(t)}$, where
$\overline{I_{\uparrow,\downarrow}(t)} =
\overline{G_{\uparrow,\downarrow}(t)\, V(t)}$ (here the overline
denotes time average). Notice, however, that while rectification would
make $I_s(B_{\rm perp}) = I_s(-B_{\rm perp})$, a quantum pumped spin
current does not need to satisfy this symmetry requirement. Thus, when
both mechanism are present, the quantum pumping component can be
partially separated by extracting the symmetric part of
$I_s(B)$. Another distinct feature of pumping is that it causes spin
transfer without voltage drop.

Recently, it was suggested \cite{moskalets01,cremers01} that while
parametric pumping does not survive the loss of phase coherence,
another mechanism of charge transfer comes into play when dephasing is
strong. We believe, however, that this incoherent mechanism, similar
to rectification, does not sustain a spin polarized current. The
reasoning goes as follows. Charge dephasing affects both quantum
pumping and rectification mechanisms for generating dc spin polarized
currents. In both cases, dephasing washes out the intricate wave
function interference patterns responsible for fluctuations in the
conductance. Even if the dephasing rate $\tau_\phi^{-1} < E_Z$, the
wave function content of spin up and spin down transport matrix
elements will become essentially the same. In that case, we expect
$I_\uparrow \approx I_\downarrow$ and therefore no net spin current.


We thank J. A. Folk and C. Lewenkopf for useful discussions. This work
was supported in part by the Alfred P. Sloan Foundation (C.C.), NSF
Grants DMR-98-76208 (C.C.) and DMR-00-72777 (C.M.M.), and by the ARO
MURI DAAD19-99-1-0215 (C.M.M.). E.R.M. acknowledges financial support
from the Brazilian agencies CNPq, FAPERJ, and PRONEX.


\end{multicols} 


\begin{references}
\bibitem{obs0} Notice that the enhancement by a factor of 2 has
nothing to do with any change from orthogonal to unitary symmetry
class.
\bibitem{obs1} We assume that a weak, nonquantizing perpendicular
magnetic field is also applied and time-reversal symmetry is broken at
all times. We thus carry our calculations for the unitary symmetry
class.
\bibitem{obs2} By restricting our analysis to this limit, we avoid
having to consider heating and nonequilibrium effects. For example,
in Ref. \cite{switkes99}, $\omega/2\pi \approx 50$ MHz $\approx 0.11
\mu$eV.
\end{references}
\end{document}